\documentclass[10pt,a4paper]{article}

\usepackage[utf8]{inputenc}
\usepackage[T1]{fontenc}
\usepackage{microtype}
\usepackage[sc]{mathpazo}
\usepackage{rotating}

\usepackage[dvipsnames]{xcolor}
\usepackage[hyphens]{url}
\usepackage{hyperref}
\hypersetup{
    colorlinks=true,
    allcolors=NavyBlue
}

\usepackage{graphicx}

\usepackage[bottom, hang]{footmisc}
\usepackage[
    top=2.5cm,
    bottom=2.5cm,
    left=2cm,
    right=2cm,
    footskip=1cm,
    headsep=0.75cm,
    columnsep=20pt,
]{geometry}

\usepackage{fancyhdr}
\patchcmd{\maketitle}{plain}{empty}{}{}

\usepackage{natbib}
\setcitestyle{numbers,square,comma}
\usepackage{titling}
\pretitle{\begin{center}\huge\bfseries}
\posttitle{\end{center}}
\patchcmd{\maketitle}{plain}{empty}{}{}

\usepackage{booktabs}

\usepackage{caption}
\usepackage{enumitem}
\setlist{noitemsep}

\usepackage{hyperref}

\usepackage{amsmath,amssymb}

\title{What Would GPT Click: Practical Effects of Human-AI Behavioral Misalignment and the Cost of Synthetic Participants in User Experience}
\author{Eduard Kuric\textsuperscript{1,2}\thanks{Corresponding author: \href{mailto:eduard.kuric@stuba.sk}{eduard.kuric@stuba.sk}\\ORCID(s): 0000-0002-7371-5512 (E. Kuric), 0000-0002-4111-1052 (P. Demcak), 0000-0001-9030-7337 (M. Krajcovic)\\}, Peter Demcak\textsuperscript{2} and Matus Krajcovic\textsuperscript{1,2}
}
\date{
\footnotesize\textsuperscript{1} Faculty of Informatics and Information Technologies, Slovak University of Technology, Ilkovicova 2, Bratislava, 84216, Slovakia
\\ \textsuperscript{2}
UXtweak Research, UXtweak j.s.a., Cajakova 18, Bratislava, 81105, Slovakia\\
}

\begin{document}

\maketitle

\begin{center}
\normalfont\bfseries\vspace{0.5\baselineskip} \abstractname
\end{center}
\begin{quote}
\normalfont\small
Synthetic participants represent a methodologically concerning concept that threatens the integrity of UX research. Findings from previous experiments specify how AI outputs are misaligned with the behaviors and thoughts of real humans in various ways. However, industry voices keep underestimating their severity, advocating for practical compromises where good-enough data does not need to be perfect, and all issues will be solved by future tuning. Our study tackles the lack of systematic understanding of the practical issues that arise with synthetic behavior and its use for steering decisions within real contexts. Within twelve diverse first click tests (n = 3431) obtained from real UX practice, we examine the ability of GPT to predict where humans click and how they reason about their behavior. Results (e.g., significantly different distribution from real data in 53\% of tasks) demonstrate critical failures to reflect the patterns in which users click on visual elements and the underlying cognitive processes. Participant personas, chain-of-thought reasoning in GPT, and different sampling parameters fail to create sensible fidelity improvements apart from inflating believability. We expose a multitude of nuanced distortions in synthetic responses that reduce their overall analytical usefulness as a decision-making resource, compared with real data. Observed distortions can be theoretically linked to the properties categorically inherent to LLMs: their statistical nature and encoding of semantic heuristics dependent on their training on linguistic data.
\end{quote}

\begin{quote}
\end{quote}

\section{Introduction}

The capacity of generative AI (GenAI) models — particularly LLMs — to participate in psychological, social and UX research has been the subject of hot discourse and controversy \citep{guest2025}. Recent headlines featured Aaru, a startup whose multi-tier fundraising tiers soared to \$1 billion, boldly proclaiming itself a competitor to research. Their pitch revolves around synthetic participants, ostensibly capable of providing perfect predictions about how people will think or behave in any situation. Experts may easily dismiss this as overhype, given that LLMs as a source of insights violate essentially every type of validity (internal, external, construct and conclusion). Mechanistically, they engage in non-transparent stochastic reproduction of the semantics from existing texts, regressively hallucinating meaning from any context available to them without actually understanding it. While this may appear as sufficient grounds for categorical rejection of synthetic participants (similarly to the declaration for AI in qualitative analysis by \citet{jowsey2025}), investigating the evidence can be posited as a more productive and scientifically curious approach.

If synthetic participants could accurately predict the nuances of human behavior and psychology, empirical evidence should support this. Existing studies indicate the opposite. Although \citet{sanders2023} showed high correlations between real and synthetic data in survey questions where answers can be inferred from past surveys, they also found that the models were unable to extrapolate to brand new polled information (support for war in Ukraine). Studies on controlled behavior have shown a lack of human-like reasoning during puzzle-solving \citep{shojaee2025}, as well as discrepancies from human strategies employed in an economic 11-20 game \citep{gao2025}. Sensory stimuli were found to evoke associations and emotional reactions in humans that LLMs did not capture \citep{imschloss2025}.

Our article addresses the lack of generalizable and interpretable behavioral evaluation of synthetic participants within UX research, The most closely related studies have limited ecological and external validity \citep{gao2025, imschloss2025, binz2023, kovac2024}. Proponents of synthetic participants might argue about their over-focus on behavior in niche psychological experiments, conceptually divorced from typical user behaviors in contexts more likely encoded in the training data (e.g., reviews and feedback, posts and comments describing user habits, decisions and experiences, published reports from UX studies, images of graphical user interface designs). Hypothetically, LLMs might perform better in practical scenarios that UX researchers, product owners and designers are interested in exploring. Without evaluation and interpretation of impact across multiple studies that focus on a specific UX method used in practice, human-AI discrepancies might be dismissed as “too theoretical”, vague and insignificant. Statistical similarity might be misinterpreted to weigh synthetic results as good-enough \citep{guest2025, sanders2023}. Additionally, the goalposts might continually shift when AI fails to meet expectations, asserting that different configurations, prompting methods (e.g., more detailed personas), or future models will solve these problems.

First click testing (FCT) — the method at the center of our research — is a more controlled variant of usability testing \citep{tomlin2018}. FCT enables standardized evaluation of clicks within visual stimuli by restricting other types of user behavior. As a rapid method that only requires an image to run, FCT has practical applications throughout the design process, from evaluation of early wireframes to final screenshots. These factors make FCT into an ideal method for an ecologically valid examination of the behavioral fidelity of multimodal LLMs. Our aim is not only to assess the fidelity with which LLMs can represent real click behavior, but also to obtain specific qualitative understanding of how and why the click behavior is misrepresented. Our findings illuminate the impact that synthetic behavioral simulation can have in practice.

Correspondingly, our study makes the following empirical and theoretical contributions to the current state of knowledge:

\begin{itemize}
    \item Comprehensive large-scale analysis of twelve (12) FCT studies with 3,431 participants, obtained from real UX research practice within different organizations.
    \item Quantitative validation of significant differences between human and synthetic click behavior within real visual design stimuli, including web and mobile app GUIs and graphs.
    \item In-depth thematic analysis of patterns differentiating human and synthetic behavior and reasoning, including critical issues of misalignment between the cognitive processing of stimuli and LLMs’ semantic associations and errant incorporation of language pragmatics.
    \item Cross-model replication with GPT-4.1 and GPT-5.2. Results indicate that different sampling methods, prompting with different types of personas, and chain-of-thought reasoning have superficial effects that do not turn LLM participants more humanlike.
\end{itemize}

Given the inaccuracies in behaviorally simple FCT, we can firmly conclude that LLMs also cannot be relied on to predict even more complex user behaviors in conditions involving more confounding variables, such as usability testing of multi-step user flows.

The structure of this article is as follows. \hyperref[sec:related]{Section 2} elaborates on key background concepts and relevant previous research. \hyperref[sec:method]{Section 3} presents our methodology and study design decisions. \hyperref[sec:results]{Section 4} presents the findings of our hybrid analysis. \hyperref[sec:discussion]{Section 5} contains the discussion that interprets our findings, arriving at theoretical and practical implications, and identifying their limitations. \hyperref[sec:conclusion]{Section 6} contains the conclusion.

\section{Background and related work}
\label{sec:related}

In this section, we methodologically situate first click testing based on its role within UX research. We review automation methods from past studies to illustrate the potentials and challenges that could render LLMs as promising in participant simulation tools and alternative sources of insight. Then, we confront such aspirations with the current state of empirical evidence that suggests key technological limitations. With this background, we highlight the gaps in understanding how LLMs behave and respond to questions in first click tasks in comparison to actual humans, which our research aims to address to support future methodology and theory building.

\subsection{First Click Testing}

In graphical user interfaces (GUIs), clicks and taps are the most common type of user behavior. They provide a trace of the intended actions of the user, making them integral in usability testing to reveal user expectations, mistakes, and even frustration \citep{kuric2025hotspots, sanchez-rola2020}. On a page or a screen, the first clicks typically indicate decisions made by users in response to the GUI stimulus (e.g., a search engine results page) to complete a task \citep{barry2011}. 

First click testing (FCT) is a technique that involves asking participants where they would click in a static visual stimulus — an image input such as a wireframe, a mockup or a screenshot — to solve a particular task \citep{tomlin2018}. FCT evolved as a specialized form of usability testing dedicated to evaluating individual designs instead of user flows. Its outputs are highly interpretable, with granular details obtainable for any image. This facilitates easy implementation even during the early design stages, as well as rapid iteration \citep{tomlin2018}. Outputs of FCT comprise the clicked locations (analyzed through heatmaps and areas of interest) and completion time as an indicator of cognitive load. They can be accompanied by gaze tracking to juxtapose clicks with visual attention \citep{chinal2017}.

\subsection{Synthetic Participants and Automated Click Evaluation}

Modeling and prediction of click/tap behavior are long-standing pursuits in information retrieval and human-computer interaction \citep{malkevich2017, do2021, chuklin2022}. However, due to the challenge of simulating the cognitive processes that underlie click behavior, prior methods were limited to predicting click propensity during search based on models of the user’s query and click history \citep{malkevich2017, chuklin2022}, or simulated cursor movements based on the positions of the cursor and the click target on the screen \citep{do2021}. Using deep learning, \citet{schoop2022} predicted affordance — whether GUI elements are perceived as tappable.

In the absence of methods dedicated to FCT automation, we analyzed similar approaches for Usability Testing as the parent method. Most studies investigating automation of usability testing focused on the automation of processes associated with its administration and data analysis rather than fully automated simulation of the FCT activity itself \citep{villamane2024}. Some methods involved detection of usability issues based on participants’ emotions \citep{esposito2022, filho2015}. Acoustic, visual and transcript features were extracted from think-aloud protocol recordings and video by \citet{fan2019, soure2021} to provide practitioners with augmentation tools for finding usability issue encounters in testing sessions. \citet{batch2023} expanded this to the visualization of behavioral features from parallel video and audio streams. However, these methods remained centered on the analysis of human data obtained within the contextually-valid tested conditions.

The advent of LLMs led to the first attempts at full automation of usability testing. The high flexibility, adaptability and linguistic capability supported constructed simulacra of cognitive and behavioral interactions. The tool SimUser by \citet{xiang2024} used LLM agents with chain-of-thought reasoning to simulate app interactions of different user groups, demonstrating mixed results with the coverage of heuristic usability issues and a lack of nuance in representing characteristics of groups. The UXAgent with a memory stream and a two-loop reasoning system was designed by \citet{lu2025} for piloting and iterating usability testing studies before involving real participants. Expert evaluation of the system showed mixed scores for accuracy, trustworthiness and insightfulness. \citet{liu2024} implemented an LLM-driven tool for automated GUI testing that outperforms the capacity of baseline solutions to find bugs.

Consequently, these studies do not provide clear granular insights about the patterns that manifest in LLM-simulated click behavior within visual designs or how they generalize across different website or app GUIs. Within this state of knowledge, we situate our study as a solution to this knowledge gap. We focus on FCT because of its practicality and simplicity, which facilitates complex evaluation and behavioral assessment in individual stimuli. Its simplicity could also make FCT into a suitable benchmark before evaluation in more complex behavioral simulation scenarios.

\subsection{Vision-Language Models in Behavioral Research}

For the prediction of FCT results, image stimuli that are visually perceived by humans represent the central inputs of the evaluated solution. As such, we analyzed the capabilities of Vision-Language Models (VLMs) in modeling human visual processing. VLMs are multimodal LLMs that incorporate computer vision algorithms with origins in image captioning to obtain descriptions of image inputs, such as GUI screenshots \citep{zang2024, wei2025}. Images are preferable to document structure information, which can be misattuned to the visual hierarchy, too cumbersomely complex to process, or unavailable \citep{cheng2024}. However, Vision-Language Models can struggle with vision tasks other than captioning, with problematic tasks including detection of the bounding boxes of objects \citep{zang2024}. This can present a challenge when designing agents that perform tasks within GUIs \citep{cheng2024}.

Importantly for simulation of cognitive processes, \citet{imschloss2025} found that although LLMs can accurately caption image contents, they can fail to make inferences and interpretations comparable to humans. The primary exceptions consisted of specific associations that had a high likelihood of being highly represented in the training datasets. This includes recall of the Kiki-Bouba effect by name when applying it during its evaluation, or attributing a calming effect to the scent of lavender. A similar phenomenon might rationally be observed in the inference of the latent meanings of GUI stimuli.

\section{Study method}
\label{sec:method}

The primary aim of our study was to examine the algorithmic fidelity of LLM-driven simulations of first click behavior against the ground truth of real user behavior. As such, the independent variable was represented by the categorical distinction between real data and simulation. For a comprehensive evaluation, the dependent variables were two-fold:
\begin{itemize}
    \item click target data,
    \item self-reported measures collected by researchers to enhance the understanding of first click behavior, typically pertaining to reasoning (e.g., justifications, evaluated constructs such as certainty or task difficulty) and the surrounding context (e.g., impressions).
\end{itemize}

Reflecting this, we sought to answer the following research questions:\\

\phantomsection\label{sec:rq1}
\emph{RQ1: How accurately do first click decisions of LLMs match human behavior?}\\

\phantomsection\label{sec:rq2}
\emph{RQ2: How accurately do LLM responses to follow-up questions in first click testing match human responses?}\\

Our secondary aim was to thoroughly investigate the degree and manner in which the various design factors of the LLM-based simulation may meaningfully improve its fidelity. Firstly, given the common expectation that future models will bring cognitive improvements \citep{binz2023} although the reliability of larger models in representing intelligence has been disputed \citep{zhou2024}, we compared the baseline to a more advanced LLM than implements iterative chain-of-thought (CoT) reasoning (for details, see Participants and materials). Secondly, since LLMs can struggle to match the variability found in human data \citep{liu2025}, we also investigated the effects of sampling parameters (temperature and nucleus/top-p) to steer the model to higher diversity by increasing the randomness of the generated output. 

Finally, persona simulation has been suggested as a promising approach that involves simulating individuals (single-persona), groups (multi-persona) or populations (mega-persona) by supplying the model with their specification \citep{gerosa2024}. Stereotypicality, lack of depth and ineffectiveness in capturing latent characteristics found in LLM responses by previous studies \citep{sanders2023, jiang2024, giorgi2024} might emerge and reduce the behavioral fidelity in FCT if the tasks evoke non-trivial cognitive complexity. Nonetheless, it is interesting to study how providing detailed personas to LLMs in an attempt to reduce the distance between real and textually simulated behavior for a purer study of the simulation effect (lack of genuine perception, reasoning and memories) impacts the outcomes.

We treat the above factors as moderating variables in order to answer the following research questions:\\

\phantomsection\label{sec:rq3}
\emph{RQ3: How does a newer LLM with CoT reasoning affect the accuracy of LLM-generated first clicks?}\\

\phantomsection\label{sec:rq4}
\emph{RQ4: How does temperature sampling affect the accuracy of LLM-generated first clicks?}\\

\phantomsection\label{sec:rq5}
\emph{RQ5: How does nucleus sampling affect the accuracy of LLM-generated first clicks?}\\

\phantomsection\label{sec:rq6}
\emph{RQ6: How does persona cardinality (individual vs. population) affect the accuracy of LLM-generated first clicks?}\\

\phantomsection\label{sec:rq7}
\emph{RQ7: How does persona specificity (generic vs. real sample definition) affect the accuracy of LLM-generated first clicks?}\\

Pre-training and fine-tuning models with real data may also improve their alignment. However, we do not explore these because of their dependence on existing data. Our focus is on scenarios where research data is unavailable and collecting it invalidates the purpose of simulations, i.e., to obtain data. Additionally, while LLMs may align with data they were trained or tuned with and make statistically accurate predictions, this may not generalize to novel observations or patterns that cannot be interpolated from past data \citep{sanders2023, brand2024}.

\subsection{Participants and materials}

The methodological aims of our study included high generalizability and ecological validity to gain understanding of LLM-simulated click behavior with close proximity to real design practice. Depending on the study domain, goals, organizational practices, constraints and other factors influenced by real-world complexity, FCT studies can be highly heterogeneous in their design. Therefore, as the source of real first click tasks, stimuli as well as behavioral benchmarks from real participants, we obtained twelve (12) studies from different organizations. The studies originate from the real practice of users of the UXtweak research platform\footnote{UXtweak user research platform: \url{https://www.uxtweak.com/}}. Data owners provided informed consent for access to their study data, exclusively for the purposes of this research, protecting their user privacy. Collectively, the obtained dataset contains \textit{n} = 3,431 participants, 45 tasks and 120 follow-up questions (30 single-choice format, 48 Likert scales and 45 open-ended questions). Designs evaluated by the studies cover various digital products and services, with domains including healthcare, public administration, social services, entertainment, and transportation. Nine studies focused on desktop versions of websites, two on mobile apps and one on diagrams. The tasks involved, among others, purchasing products and activating memberships, locating context-specific information and navigating content.

The studies followed a standard structure with customizable steps. Their core comprised a welcome message, instructions and multiple first click tasks. Optionally, based on user needs, studies could include a pre-study questionnaire and follow-up questionnaires after individual tasks and at the study end. The acquired results included task solutions (clicks locations within stimuli), descriptive data of the samples (answers from pre-study questionnaires, screening questions, targeting attributes from panel recruitment configurations), and follow-up question responses.

As the baseline LLM model, we opted for GPT-4.1, a non-reasoning model that improved upon the capacity of its predecessors to process long context and follow instructions\footnote{OpenAI API documentation - GPT-4.1: \url{https://developers.openai.com/api/docs/models/gpt-4.1}}. For \hyperref[sec:rq3]{RQ3}, we used GPT-5.2, OpenAI’s latest model available at the time of the experiment. GPT-5.2 implements CoT reasoning and is correspondingly advertised for its ostensibly highest reasoning capacity\footnote{OpenAI API documentation - GPT 5.2: \url{https://developers.openai.com/api/docs/models/gpt-5.2}}. Although alternative models (e.g., Claude, LLaMa, Gemini) may achieve moderately diverse performances in specific tasks because of differences in training and data, past studies have found shared issues in algorithmic fidelity in various contexts \citep{liu2025, almeida2024}. As such, we adopt the above OpenAI models as representatives of the paradigm, reasonably predicting that other LLMs would exhibit nonidentical but similar behavior.

\subsection{Procedure}

Our experimental procedure involved conversational prompting of LLMs to re-enact the twelve first click studies from our dataset. For simulated replications to be as faithful as is within the capacity of the model, models were provided with the same information that was available to the participants. For comprehensive evaluation of the research questions, each study was simulated across six configurations. These configurations, shown in \autoref{tab:conditions}, correspond with the baseline condition and the variables evaluated as moderators. The solution that can be conventionally predicted as the best comprises the baseline (\texttt{temperature = 1} and \texttt{top\_p = 1} for varied yet coherent output \citep{ma2025b}, mega-personas to approximate population diversity \citep{gerosa2024}, and specific personas to match the description of the original sample as context). As a single exception, the baseline implements GPT-4.1 over GPT-5.2 since the latter is incompatible with sampling parameters in its reasoning mode.

As proprietary models, background technical information that differentiates GPT-4.1 and GPT-5.2 (e.g., fine-tuning methods, number of parameters) is not publicly available. Although CoT reasoning presents a key value proposition of GPT-5.2 based on its advertising, to isolate its effect, we evaluated GPT-5.2 twice. First, we set the reasoning effort parameter to none and then to medium.

Evaluations were originally planned to aggregate multiple runs within the same conditions, considering the nondeterminism of LLMs that can result in considerably different outputs. However, despite the generated text being nominally different, pretests with the analyzed measures across three runs showed minimal differences in observed metrics in all twelve studies. This aligns with previous studies identifying hyperaccuracy and lack of variability in LLM outputs \citep{liu2025, amirova2024}. Therefore, due to the complexity and scale of our methodology (84 simulations total), we made the computationally-efficient and environmentally-conscious decision to compare single runs.

\begin{table}[!ht]
\caption{Simulation conditions related to research questions.}
\label{tab:conditions}
\centering
\begin{tabular}{lcccccc}
\toprule
\textbf{RQ} & \textbf{Model} & \textbf{temperature} & \textbf{top\_p} & \textbf{Persona Type} & \textbf{Persona Spec.} & \textbf{Reasoning} \\
\midrule
\hyperref[sec:rq1]{RQ1}, \hyperref[sec:rq2]{RQ2} & gpt-4.1 & 1 & 1 & mega & specific & inapplicable \\
\hyperref[sec:rq3]{RQ3} & \textbf{gpt-5.2} & 1 & 1 & mega & specific & \textbf{none} \\
\hyperref[sec:rq3]{RQ3} & \textbf{gpt-5.2} & inapplicable & inapplicable & mega & specific & \textbf{medium} \\
\hyperref[sec:rq4]{RQ4} & gpt-4.1 & \textbf{0.2} & 1 & mega & specific & inapplicable \\
\hyperref[sec:rq5]{RQ5} & gpt-4.1 & 1 & \textbf{0.2} & mega & specific & inapplicable \\
\hyperref[sec:rq6]{RQ6} & gpt-4.1 & 1 & 1 & \textbf{single} & specific & inapplicable \\
\hyperref[sec:rq7]{RQ7} & gpt-4.1 & 1 & 1 & mega & \textbf{nondescript} & inapplicable \\
\bottomrule
\end{tabular}
\end{table}

\begin{figure}[!ht]
    \centering
    \includegraphics[width=\linewidth]{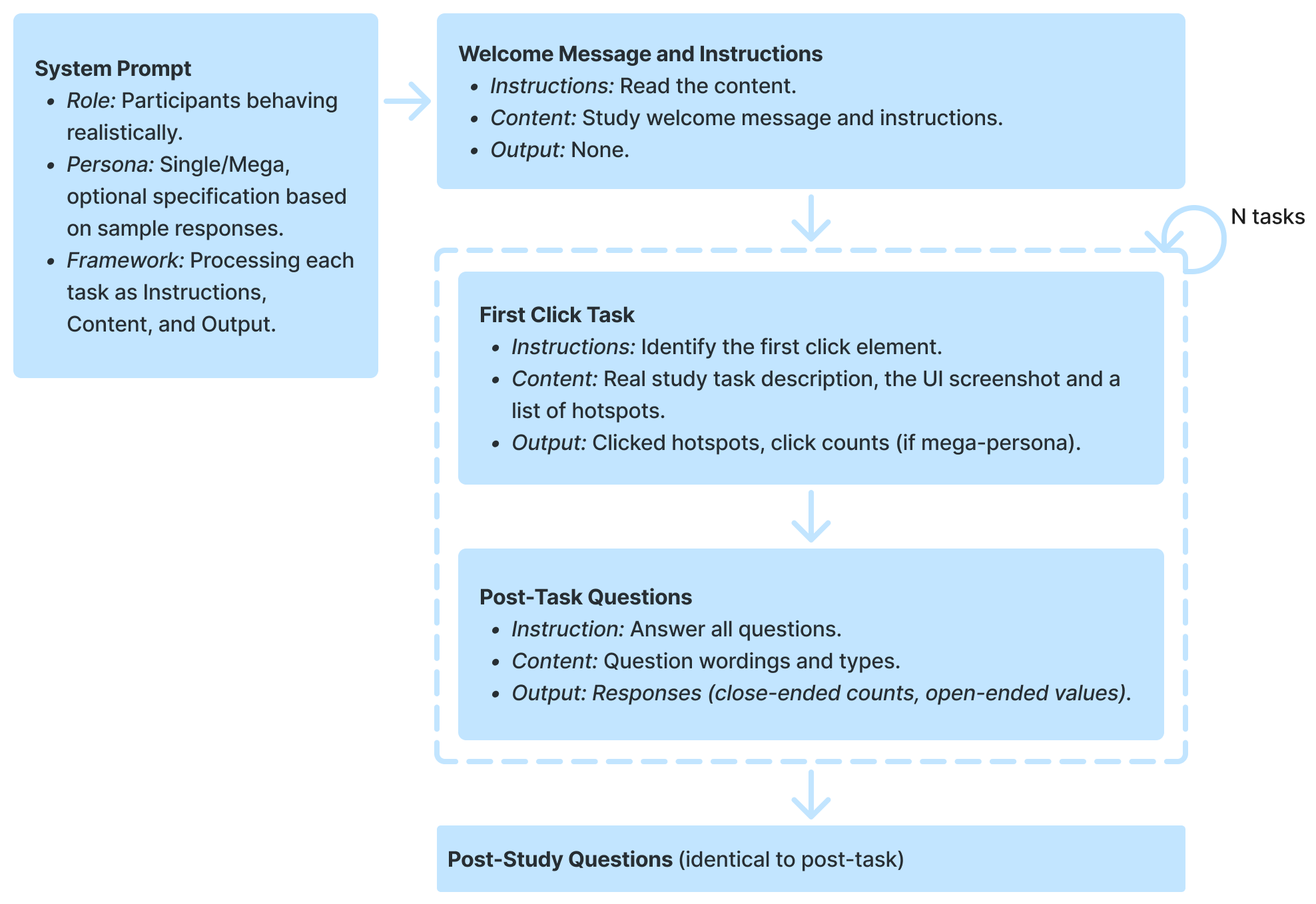}
    \caption{Simulation procedure flowchart. Instructions are meta-directives applied across studies, content represents study-specific instructions and stimuli from the original source, output defines format.}
    \label{fig:procedure}
\end{figure}

The simulation procedure, shown in \autoref{fig:procedure} consists of prompts that guide the LLM through the simulated FCT activity. During the First Click Task step, multimodal prompts included the original image-based stimuli, as well as a listed hotspots - clickable areas that serve as valid answers. In accord with \citep{cheng2024} and \citep{zang2024}, our pretests revealed a failure of LLMs to locate elements in stimuli with accurate coordinates. To address this challenge within the capacity of the evaluated models, we used hotspots (areas of interest) rather than pixel coordinates as the answer format. To avoid ambiguity, hotspots were defined by their label, descriptive types and location (e.g., Sign in menu item in the upper right). Hotspots were defined for all interface elements that were either clicked by real participants, or had comparable saliency in the visual hierarchy. For new tasks, images that were no longer relevant were removed from the execution context to optimize performance. They were returned there during the post-study questionnaire to provide models with full context.

We tuned the prompts iteratively to ensure the instructions were followed precisely and outputs matched the specification (see the \hyperref[sec:data-statement]{Data availability statement} for the final prompts). This includes repetition of instructions and output formats between tasks and questionnaires. Although removal of repetition could reduce the number of input tokens, GPT tended to forget to follow key directives within the instructions without it. Differences between single-persona and mega-persona prompts match the definitions of the approaches — synthetic perspectives of individuals and results that represent an aggregated population respectively \citep{gerosa2024}.

Persona specifications matching the description of the original samples were obtained by leveraging the responses from pre-study questionnaires. They comprised general information about age, gender, education, and Big Five personality profiles, as well as specific information, including knowledge and past experiences regarding context-relevant topics. Mega-personas were described as distributions. Single-personas contained exact individual profiles. In studies where age, gender, education or personality information about participants was unavailable, personas were enriched to match distributions found in the general population to simulate random sampling\footnote{Statista - Age distribution of internet users: \url{https://www.statista.com/statistics/272365/age-distribution-of-internet-users-worldwide}}\footnote{OECD - Education attainment: \url{https://www.oecd.org/en/topics/education-attainment.html}} \citep{fleeson2010}.

\subsection{Measures and analysis}

For a robust evaluation of the algorithmic fidelity of LLM-simulated first click behavior that is both statistical and nuanced, we applied a hybrid quantitative and qualitative approach. For the quantitative portion, the measures summarized in \autoref{tab:measures} were divided between the evaluation of choices (for hotspot clicks and close-ended follow-up questions) and text (for open-ended follow-up questions). Chi-squared and Mann-Whitney tests were used to account for non-normal data distributions.

We conducted the explorative qualitative portion of our analysis with the aim of explaining the identified patterns, as well as studying the additional properties of synthetic data that the quantitative perspective might miss. With our goal being organic theme generation \citep{braun2020}, we based our method on the reflexive thematic analysis framework by \citep{braun2008}. Codes were refined iteratively to inductively fit the observations. To build themes that enable the comparison of synthetic and real data, the codes focused on patterns that can apply generally across different domains, studies and questions, such as “contextually nonsensical statement” or “generic property-focused feedback without experiential justification” (e.g., praising intuitiveness without explaining how or why the subject is intuitive). Naturally, some codes were more or less relevant depending on the context. During theme definition, we reflected on patterns and their relationships between the synthetic and real data. Concurrently, we reflected on their alignment with established theory about the algorithmic fidelity of synthetic participant responses.

\begin{table}[!ht]
\caption{Measures for comparison of synthetic and real data and evaluation of solution design factors as moderators.}
\label{tab:measures}
\begin{tabular}{lp{12cm}}
\toprule
\textbf{Name} & \textbf{Description} \\
\midrule
\multicolumn{2}{l}{\textit{Choice measures (clicks and close-ended follow-up questions)}} \\
\midrule
\begin{tabular}[t]{@{}l@{}}\textbf{Distribution Difference}\\ Values: true/false\end{tabular} & Goodness of fit, calculated as a chi-square test applied to the distribution of hotspot/choice selections to establish whether there is a significant difference. \\
\begin{tabular}[t]{@{}l@{}}\textbf{First-choice Agreement}\\ Values: true/false\end{tabular} & Simple and easily interpretable binary evaluation of whether the top choice (the hotspot clicked or answer selected by most real or synthetic individuals) matches between conditions. \\
\begin{tabular}[t]{@{}l@{}}\textbf{Rank-order Agreement}\\ Values: 0-1\end{tabular} & Choices are ranked by frequency to calculate the ratio of choices that are placed in identical order between conditions. \\
\begin{tabular}[t]{@{}l@{}}\textbf{Jensen-Shannon Distance}\\ Values: 0-1\end{tabular} & A measure for comparing probability distributions with sensitivity to order changes. Accounts for diverse values of N within the original studies. \\
\begin{tabular}[t]{@{}l@{}}\textbf{Normalized Entropy}\\ Values: 0-1\end{tabular} & A variability measure to capture the range between universal consent where all choices congregate around a single option (0) and complete randomness where all choices are represented equally (1). \\
\begin{tabular}[t]{@{}l@{}}\textbf{Unique Selection Count}\\ Values: Integer\end{tabular} & The sum of unique options that were selected at least once. Total number is used instead of a ratio to interpret variability in contexts where stimuli contain numerous hotspots, many of which may be less relevant based on participant behavior. \\
\midrule
\multicolumn{2}{l}{\textit{Text measures (open-ended follow-up questions)}} \\
\midrule
\begin{tabular}[t]{@{}l@{}}\textbf{Lexical Similarity}\\ Values: 0-1\end{tabular} & Linguistic property for comparing the words being used. Calculated in a Term Frequency-Inverse Document Frequency (TF-IDF) space \citep{lumintu2023} shared between the conditions. Its value is the cosine similarity between centroid vectors. \\
\begin{tabular}[t]{@{}l@{}}\textbf{Semantic Similarity}\\ Values: 0-1\end{tabular} & Linguistic property for comparing the meaning of sentences. The all-MiniLM-L6-v2 model, a lightweight open-source alternative to BERT transformers, was used to obtain sentence embeddings. Calculated as cosine similarity between their centroids. \\
\begin{tabular}[t]{@{}l@{}}\textbf{Semantic Diversity}\\ Values: 0-1\end{tabular} & Sentence embeddings from all-MiniLM-L6-v2 were also used to assess the overall within-condition variability of sentences. Calculated as the mean of pairwise cosine distances between sentences \citep{venkit2025}. \\
\begin{tabular}[t]{@{}l@{}}\textbf{Lexical Repetitiveness}\\ Values: Real number\end{tabular} & Repetition of words, gauged as Yule’s K \citep{hohne2024}. \\
\begin{tabular}[t]{@{}l@{}}\textbf{Readability}\\ Values: Real number\end{tabular} & An objective quantifier of reading ease calculated with Flesch’s formula for reading ease \citep{hohne2024}, calculated through the Textstat library*. \\
\begin{tabular}[t]{@{}l@{}}\textbf{Word Count}\\ Values: Integer\end{tabular} & Simple count for descriptive purposes and as context for other measures. \\
\bottomrule
\addlinespace[2pt] % Adds a tiny bit of breathing room
\multicolumn{2}{@{}p{\textwidth}}{\footnotesize *Textstat python library for text statistics: \url{https://github.com/textstat/textstat}}
\end{tabular}%
\end{table}

The stimuli — representing screenshots, mockups of user interfaces and graphs — contained many hotspots as available click options, many of these were not clicked by any participants, real or synthetic. Close-ended questions also contained options that were selected only rarely or not at all. As a uniform approach for only analyzing relevant options, we only incorporated options that were selected at least once.  To meet the assumptions of chi-squared tests by avoiding average cell frequency below five, within these tests, we created a singular category to aggregate those with the average frequency below five.

\section{Results}
\label{sec:results}

This section presents the empirical observations from our study that provide answers to our research questions. Complete results of statistical analysis are summarized in \autoref{tab:stats}. For their further interpretation, see \hyperref[sec:discussion]{5. Discussion}.

\subsection{Behavioral fidelity of first clicks (RQ1)}

\emph{\hyperref[sec:rq1]{RQ1}: How accurately do first click decisions of LLMs match human behavior?}\\

\begin{figure}[!ht]
    \centering
    \includegraphics[width=\linewidth]{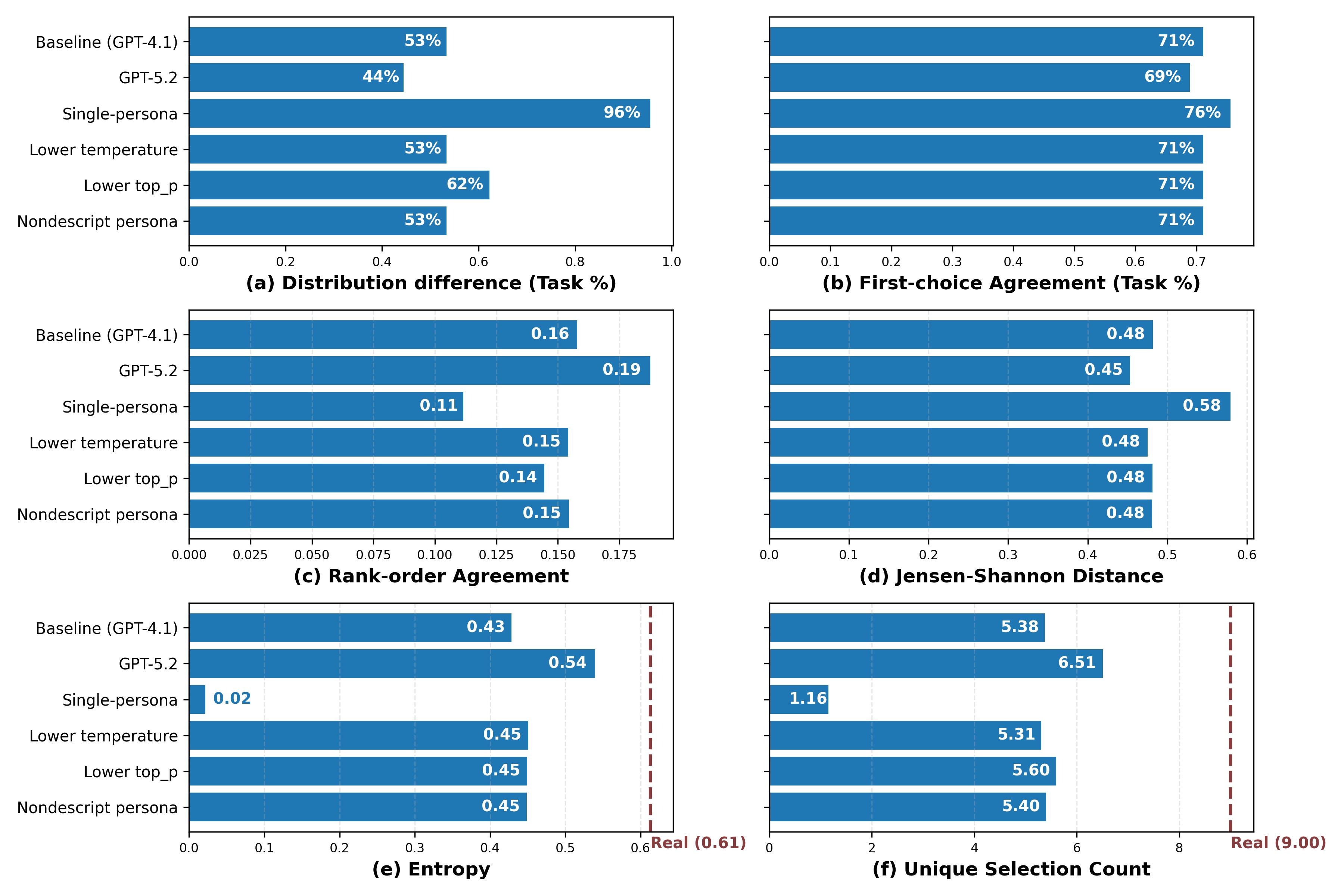}
    \caption{Comparison of first click behavioral measures between simulation conditions, evaluated using data from all studies.}
    \label{fig:click-measures}
\end{figure}

Comparative analysis of the baseline model indicates statistically significant Distribution Difference in 53\% of tasks. The hotspot clicked most frequently by GPT aligned with the human first choice in 71\% of tasks. The average Rank-order Agreement was .16 (\textit{SD} = .14). The mean Jensen-Shannon distance was .48 (\textit{SD} = .16), suggesting a moderately high level of dissimilarity present across most studies given the 0-1 range. Entropy of synthetic clicks (\textit{M} = .43, \textit{SD} = .19) was significantly smaller than of real data (\textit{M} = .61, \textit{SD} = .18), $z = -4.09$, $p < .001$, $r = -.43$ $[-.60, -.26]$. This was reflected in the number of unique selected hotspots as \textit{M} = 5.38 (\textit{SD} = 2.44) in synthetic data and \textit{M} = 9.00 (\textit{SD} = 4.75) in real data, $z = -4.35$, $p < .001$, $r = -.46$ $[-.62, -.28]$. These results are illustrated for comparison with other conditions in \autoref{fig:click-measures}.

Our qualitative analysis of clicks enabled a more nuanced understanding of the differences between simulated and real human click behavior. In most stimuli, there was a single hotspot that GPT resolved as the “victor”, to which the model attributed the overwhelming majority of clicks. This happened irrespective of whether the victor was actually the correct solution based on design intent or participant solutions. Since GPT is a blackbox, the method for selecting the victor is unknown, although we found that they were all elements with some semantic connection to the task. GPT did not apply this strategy in some extremely cluttered and cognitively complex stimuli where the relations between the difficult tasks and the presented information were obscure, although this could be attributed more to its failure to do so, given that the selected hotspots were still misaligned from real clicks. 

For comparison, similar victors were found in human data in tasks where the phrasing contained a whole or partial label from the intended solution. Guiding participants to the solution in this manner reduces the methodological validity of such tasks. For external and ecological validity, these tasks were kept in our dataset as a representation of real-world practical studies that establish how simulated first clicks might appear from the perspective of inexperienced researchers and non-research practitioners involved in UX democratization within the industry. Even in some of these guiding tasks, humans still applied higher-level problem solving, as evidenced by alternative solutions reliant on meaning, whereas GPT selected its victor based on the matching words.

Inconsistencies between real and simulated click patterns can be divided into two categories: spurious clicks and omissions. Spurious clicks projected a false popularity for hotspots that contained zero or very few clicks by real participants. A common pattern entailed clicks on unintuitive or high-level labels that require a leap in logic to link to the task. It also favored elements with more standardly interactive appearance, like buttons, links and menu items. Other examples included clicks on the search field when most people solved the task through navigation, or clicking social buttons to find information away from the website that they were already on. Both of these could hypothetically represent valid solutions in some scenarios, but real people ignored them within the stimulus and task context. Some generated clicks were also unambiguously nonsensical.

Simulations commonly omitted hotspots that featured prominently in the real results (e.g., third, second, even first). If their click count was non-zero, it was small-enough as to be indistinguishable from other marginal options, including the spurious ones, which GPT presented alongside the victor, generating an illusion of variability. Human clicks showed clear interpretable misunderstanding of labels and associations that were completely missing from simulations, such as clicks on similar labels with different meanings. Human clicks highlighted expected interactivity of banners and labels in the hero section that GPT mostly ignored. Also unlike GPT, when the stimulus failed to clearly communicate the expected information, some humans clicked the contact or download of additional materials.

Less commonly, GPT selected the same group of top solutions as humans instead of a single victor, but then assigned them a different order. This was represented mostly in cases when multiple solutions were valid, such as alternative navigation modes (e.g., a menu and links in the page content), albeit inconsistently (arbitrary victors still manifested). Tasks where identical conclusions about clicks could be drawn from the simulated results were rare, resulting from randomness and guiding task phrasing.

Collectively, these findings provide the following answer to \hyperref[sec:rq1]{RQ1}: Click decisions simulated with LLMs are inaccurate and unreliable as a predictor of human click behavior, with different order patterns and unrealistically low variability.

\subsection{Reasoning fidelity in follow-up questions (RQ2)}

\emph{\hyperref[sec:rq2]{RQ2}: How accurately do LLM responses to follow-up questions in first click testing match human responses?}\\

Our examination of follow-up answers was bidirectional, divided between close-ended and open-ended questions. Close-ended questions comprised mostly rating stimulus attributes (e.g., intuitiveness, enjoyability), feelings (e.g., confidence, comfort), asking the participants whether they noticed specific information or about their willingness to use a solution. Significantly different answer distribution was found in 45\% of these questions, with a 50\% First-choice Agreement and Rank-order Agreement of \textit{M} = .36 (\textit{SD} = .27). The Jensen-Shannon distance was \textit{M} = .28 (\textit{SD} = .13). Entropy, at \textit{M} = .76 (\textit{SD} = .08) was comparable to its real data counterpart at \textit{M} = .72 (\textit{SD} = .15), $z = 0.75$, $p = .45$. The difference between Unique Selection Counts was close to statistical significance, lower and less varied for synthetic responses (\textit{M} = 4.86, \textit{SD} = 1.31) than in real responses (\textit{M} =  5.15, \textit{SD} = 1.67), $z = -1.62$, $p = .094$.

\begin{figure}[!ht]
    \centering
    \includegraphics[width=\linewidth]{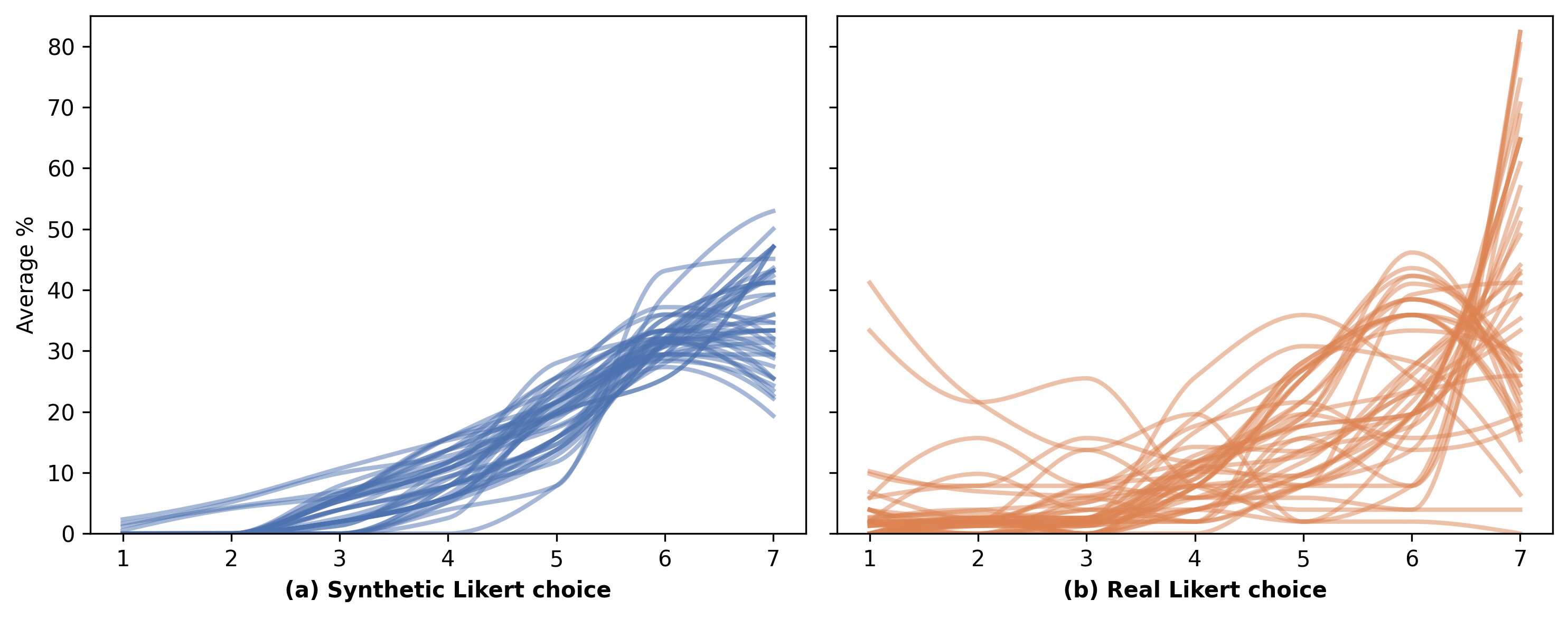}
    \caption{Line chart of Likert scale response distributions plotting all tasks to compare synthetic data (left) and real data (right). Statistically, score 7 synthetic \textit{SD} = 8.68\%, real \textit{SD} = 22.58\%; score 1 synthetic \textit{SD} = 0.47\%, real \textit{SD} = 7.98\%.}
    \label{fig:likert-diversity}
\end{figure}

Qualitatively, the above statistical differences manifested in single-choice questions in contrasting ways. In multiple instances, simulations yielded  opposite answers (e.g., overestimating whether people saw a piece of information) and overly balanced magnitudes (e.g., underestimating the willingness to recommend a product). While Likert scales showed a positivity bias in both types of data, this description marginalizes salient distributional differences. Synthetic results were significantly less varied. As shown in \autoref{fig:likert-diversity}, they had a homogenous moderate skew with a mode peaking on the highest or second highest rating regardless of the question. Real data contained heterogeneous patterns including multimodality, even, neutral and negatively skewing distributions, along with almost perfect consensus (e.g., minimal difficulty ratings of some tasks). The simulation’s similar mean entropy with lower SD can be thus attributed to insufficiently nuanced contextual adaptability.

\begin{figure}[!ht]
    \centering
    \includegraphics[width=\linewidth]{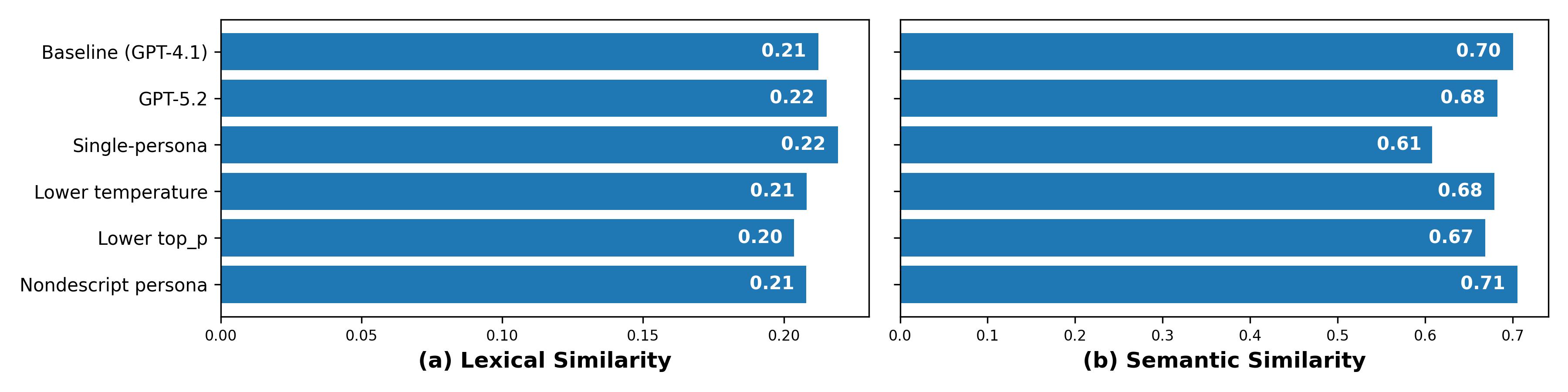}
    \caption{Linguistic similarity between synthetic responses to open-ended questions and real data across the evaluated conditions.}
    \label{fig:similarities}
\end{figure}

Open-ended follow-ups typically focused on justifications of clicks and rating choices, impressions and suggestions for the stimuli. The lexical similarity of the baseline to real responses was \textit{M} = .21 (\textit{SD} = .11) and semantic similarity was \textit{M} = .70 (\textit{SD} = .15), with comparable results across other conditions (see \autoref{fig:similarities}). For relative evaluation of these measures (since there are no generally applicable thresholds) the mutual similarity between the baseline and other LLM conditions (\hyperref[sec:rq3]{RQ3-7}) ranged between $.34-.43$ lexically and between $.86-.96$ semantically. This indicates significant differences in expression and meaning from human responses.

\begin{figure}[!ht]
    \centering
    \includegraphics[width=\linewidth]{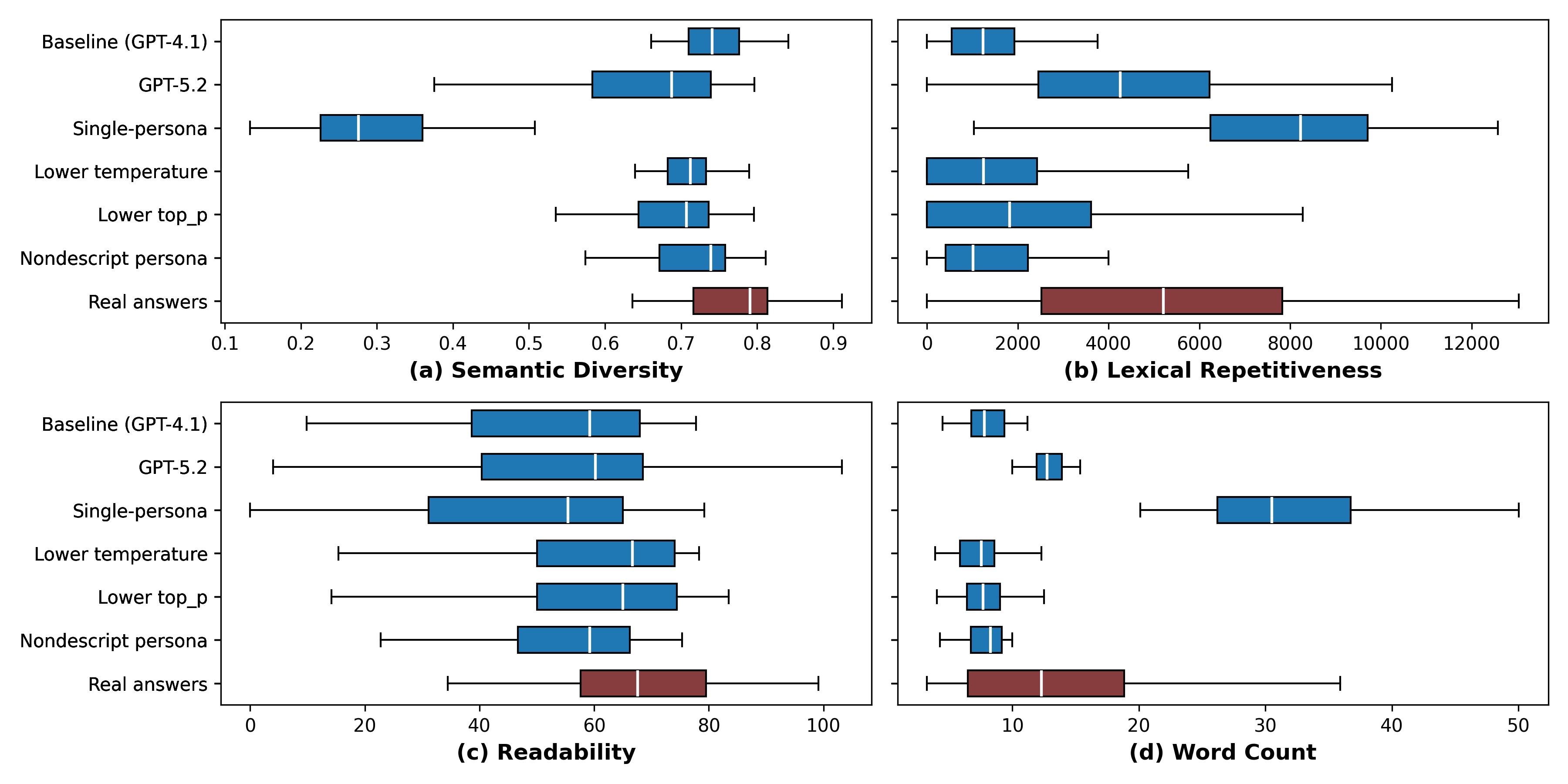}
    \caption{Boxplot comparison between synthetic and real responses to open-ended follow-up questions aggregated across FCT studies.}
    \label{fig:open-boxplots}
\end{figure}

The semantic diversity of the synthetic responses was significantly lower (\textit{M} = .70, \textit{SD} = .14) than of real responses (\textit{M} = .76, \textit{SD} = .09), $z = -2.54$, $p = .011$, $r = -.27$, $[-.46, -.07]$. Synthetic responses used words less repetitively (\textit{M} = 1579, \textit{SD} = 1652) than real ones (\textit{M} = 5606, \textit{SD} = 3943), $z = -5.62$, $p < .001$, $r = -.59$ $[-.72, -.43]$, and were also less readable (\textit{M} = 54.04, \textit{SD} = 20.76) than real responses (\textit{M} = 67.40, \textit{SD} = 16.27), $z = -3.13$, $p = .002$, $r = -.33$ $[-.51, .13]$. Human responses were longer on average at 13.31 words (\textit{SD} = 8.26) than synthetic responses at 9.08 words (\textit{SD} = 4.76), a result that is almost statistically significant ($z = -1.92$, $p = .056$). These distributional differences are illustrated in \autoref{fig:open-boxplots}.

Our reflexive thematic analysis revealed several themes that capture distortions \citep{aher2023}:

\subsubsection{Superficiality and homogeneity}

The deficient internal depth of GPT responses is an antecedent to the external lack of variability between them. The contents of GPT responses typically attributed the stimulus general traits while describing or commenting on one or multiple visual elements (e.g., \textit{“There are several navigation options, which makes it seem comprehensive.”}, \textit{“There isn’t a single named [task keyword] section, so I clicked [label]”}). Human responses also contained some comparable general statements, explaining the limited semantic and syntactic similarity to real data as the consequence of predicting likely words and their mutual relationships. The greater depth and variability of human responses stemmed from the expression of various combined expectations, observations, associations, interpretations, implications, understandings, and rationales.

This theme of superficiality can be further divided into multiple subthemes. The following patterns were found in humans responses, granting them meaningful depth that was limited or had no counterparts in simulations:
\begin{itemize}
    \item \emph{Task Factors.} Focus on how the interface helps solve the current task, or other related tasks drawing from personal experiences and expectations. Assessment and analysis of relevance and usefulness based on personal understanding of goals, wants and needs.
    \item \emph{Aesthetic \& Hedonic Factors.} Emotions that the stimulus evoked and the aesthetical impressions it left, including meaningful metaphors and similes.
    \item \emph{Cognitive \& Experiential Factors.} Some participants discussed choices they considered for an extended amount of time but did not click, or information they did not notice. Others cited personal perspectives and life experiences (e.g., \textit{“As someone who -”}), or referred to cognitive conditions and abilities when addressing specific information they found difficult to process. The closest similar feedback by GPT were generic hallucinated complaints about a small font size or a claim that it was deciding between the top choices in the simulated click data, offering no added distinction between synthetic “thoughts” and “behavior”. 
    \item \emph{Implicit Feedback.} Some human responses enabled the inference of information such as beliefs and misunderstandings, forgetting, objects of attention, word choices and persona factors that coincide with behavior. GPT responses were explicit, hyperaccurate (e.g., quoting exact labels) and devoid of layers beyond the letter of the text.
    \item \emph{Procedural Context.} Some responses stemmed from multitask interactions, such as general feedback relevant across tasks that gradually became more apparent (e.g., about navigation, visual cues, terminology), or clarifying they only knew where to look for the solution because they encountered it during a previous task.
    \item \emph{Mental Models and Lexical Nuance.} When the solution to a task was obscure, some participants listed a broader space of labels for the concepts that they were looking for. GPT typically only complained it could not find a match for an obvious label based on the task phrasing. Common synonyms for an action label were used interchangeably by GPT, yet humans showed a clear preference for one based on both explicit and implicit feedback (prevalently using a term different from the label).
    \item \emph{Socio-Cultural Change.} In one instance, LLM responses overwhelmingly complained about the semantic shift of a term being used, arguing instead for a more conservative term. Human responses commented on the term’s novelty and mentioned the same alternative as the LLM. However, they did not indicate this as a severe understanding issue and admitted the new term’s expressive advantages. This is evidence of GPT’s temporal lock-in because of rooting in past text and collation of correctness with probability.
    \item \emph{In-depth Constructive Criticism.} Exploration of a concept beyond the current state, such as spinning an idea viewed as flawed or highlighting positive aspects to pivot toward.
    \item \emph{Meaningful Tangents.} Digressions by real participants that were incongruous to the formalistic interpretation of the tasks and questions, but actually made meaningful connections and provided informative context. Valuable examples of going off-topic include expectations about design behavior, mental associations of labels, or usability issues with related processes experienced in real life.
\end{itemize}

In synthetic responses, by contrast, the following patterns embodied their limited depth:
\begin{itemize}
    \item \emph{Generic and Homogenized Topics.} Synthetic responses emphasized generics topics likely based on their high prevalence in semantically related text. Examples include nonsensical design suggestions, complaints about cognitive load, redundancy, or the language being too formal. This occurred even when the majority of human feedback focused on a specific topic that emerged as prevalent spontaneously (e.g., the fact that there was no obvious solution to the task). Even when pertaining to the same topic as human responses, synthetic responses lacked the specificity that made them interpretable and relevant (e.g., \textit{“Information is difficult to understand”} — which information and how?).
    \item \emph{Hollow Elaboration.} Prolonged responses that did not add meaningful information, unlike the examples in real data discussed above. Descriptive Commentary was a common representative of this behavior — selecting one or multiple random visual elements and essentially just confirming their existence with a superficial comment, e.g., \textit{“The [element] allows you to [element function], which is useful”}. Excuses were another type — shallow justifications for feedback, such as \textit{“based on my experience with similar apps”}.
    \item \emph{Stochastic Repetition.} Responses from different simulated participants that all paraphrased the same information (e.g., generic praises).
    \item \emph{Template-like Structure.} When simulating many responses, GPT often produced outputs with highly uniform structure despite their lexical diversity. For example, justifications of choices were often phrased as contradictions (e.g., \textit{“I [did something] but then [something else]”}).
    \item \emph{Fake Variability.} See \hyperref[sec:hallucinations]{Hallucinations} below.
\end{itemize}

\subsubsection{Hallucinations}
\label{sec:hallucinations}

It was not uncommon for GPT to generate responses with no support in human data or objective reality. Some were unambiguously nonsensical, others represented roundabout semantic associations that would require long leaps in logic. Stochastic speculation encompassed:
\begin{itemize}
    \item stimulus contents (e.g., complaints about imaginary loading speeds, not understanding a hallucinatory concept, or praise for the guidance by non-existent icons), 
    \item behaviors and cognitive processes (e.g., \textit{“Very easy once I scanned to the bottom of the page”} when all relevant content is at the top)
    \item habits (e.g., proclaiming rich experience with low-frequency or life-event-driven solutions). 
\end{itemize}

In questions with a broad space of potential answers, vast hallucinations over-compensated for a lack of depth with Fake Variability. For instance, this resulted in suggestions that, if taken at face value, would require the addition of N new features, placing everything at the top, center, more colorful, with less text, “more clear”, and following other generic heuristics. The noise obscured actual empirically supported responses, which were understated and lacking defining nuance, if not completely excluded.

\subsubsection{Positivity and Helpfulness Distortion}

Similarly to the Likert scales, open-ended responses also exaggerated the positive traits of the stimulus and its experience with high prevalence (e.g., easy, great, useful, well-organized, simple, nothing misleading, reassuring, helpful, smooth, no issues, straightforward). Criticism, if provided, was usually impersonal, general, noncommittal and mixed in with positive traits (e.g., \textit{“The layout is clear but maybe too many banners”}). It lacked the experiential rooting of human negative feedback that enables understanding and quantification of severity and effort (e.g., \textit{“[information] confused me and I had to scroll down the page thrice before I was sure there was no better choice”}). As a result, it was devoid of the rich spectrum of criticism found in empirical data, from passive dismissal (e.g., \textit{“meh, standard”}), through emotional responses to frustration, to critical failure states (e.g., \textit{“I gave up”}).

Aside from direct positivity, GPT responses also exaggerated helpfulness. In optional questions where real people replied only if they had something meaningful to say, GPT simulated a new response every time. This resulted in highly repetitive answers or overcompensated diversity where the majority of replies had no support in real data.

\subsubsection{Literal Interpretation and Over-Roleplay}

Although not always, human responses demonstrated the capacity for making pragmatic inferences about the intent behind the question and answering it accordingly. The LLM’s capacity to do the same was inconsistent to coincidental as their responses remained on the layer of meaning (semantics) while ignoring the use of language (pragmatics). This was well-represented in various instances as over-roleplay. Real participants, when given a task scenario to help them empathize with a situation, still recognized they were playing a role and the stimulus was only an inert image. LLMs — despite having the context about testing — sometimes gave responses indicative of a failure to make this distinction. This included discussing the stimulus as a product that helped them in real life, or hallucinating experience they could not have due to it being a static image.

Another critical example was found in a final question, which asked the participants whether there was any other information they would like to share. Many participants provided their feedback about the testing as a whole. GPT uniformly answered with a single word, “yes/no”, without elaborating. This is demonstrative of a lack of contextual awareness and Theory of Mind.

\subsubsection{Spurious Heuristics}

Some responses were the clear result of semantic associations that had no relevance within the context. For example, when a task included the word “family” as part of its flavor text (to help participants empathize with the scenario by framing it as their family making plans), some responses praised the design as \textit{“family-friendly”} or \textit{“great for parents”} without explanation. The stimulus was a web page with no particular family orientation or features and real participants did not label it as such. This phenomenon is demonstrative of probabilistic mechanisms by which LLMs generate their outputs by applying semantic shortcuts. The other distortions could be attributed to the same principles, albeit with more challenging traceability to specific words.

The above findings reveal the answer to \hyperref[sec:rq2]{RQ2} in the form of comprehensive and multifaceted insights into the vast differences between real and LLM–synthesized follow-up answers in FCT. The inaccuracies in synthetic responses deepen the understanding of the gap in LLM’s capacity to simulate click behavior.

\subsection{Chain-of-thought reasoning moderation (RQ3)}

\emph{\hyperref[sec:rq3]{RQ3}: How does a newer LLM with CoT reasoning affect the accuracy of LLM-generated first clicks?}\\

Our evaluation of GPT-5.2 across two reasoning settings (disabled and medium) yielded no significant differences in the clicks measures, and only limited differences in measures pertaining to follow-up questions. For a concise understanding of how CoT reasoning can and cannot improve click simulation, we compare it to the baseline directly. We highlight the differences from the non-reasoning alternative where they appear.

The introduction of GPT-5.2 with reasoning enabled into the simulation did not address the essential problems of the baseline model. Most click measures demonstrated no significant changes, as shown in \autoref{fig:click-measures}. The only change was reflected in Entropy which was significantly higher than the baseline at \textit{M} = .54, \textit{SD} = .14 ($z = -2.77$, $p = .006$, $r = -.29$ $[-.48, -.10]$) yet remains significantly lower than in the real data. Correspondingly, Unique Selection Count was also higher at \textit{M} = 6.51, \textit{SD} = 1.93 ($z=-3.49$, $p < .001$, $r = -.37$ $[-.54, -.18]$).

In close-ended follow-ups, the reasoning model was even less aligned with humans than the already deviated baseline (see \autoref{tab:stats}). This was indicated by most metrics, including Distribution Difference increased by 17\%; First-choice Agreement reduced by 12\%. Enabling CoT provided only small improvements that eliminated the statistical significance of differences from the baseline: Rank-order Agreement to \textit{M} = .30, \textit{SD} = .29 (without reasoning \textit{M} = .27, \textit{SD} = .3), Jensen-Shannon Distance to \textit{M} = .30, \textit{SD} = .14 (without reasoning \textit{M} = .32, \textit{SD} = .14), and Entropy \textit{M} = .77, \textit{SD} = .11 (without reasoning \textit{M} = .79, \textit{SD} = .10). In the textual analysis of open-ended answers, the only difference was in Semantic Diversity, where reasoning actually reduced its value from \textit{M} = .67, \textit{SD} = .13 to \textit{M} = .64, \textit{SD} = .14 (significantly lower than the baseline, $z = 3.00$, $p = .003$, $r = .3$2 $[.11, .50]$, which was still lower than in real data). 

Qualitative analysis indicated that the differences between the entropy of the baseline and reasoning models stemmed primarily from several tasks where clicks were re-shuffled in a way that made the victors less prominent. Multiple options with semantic relations to the task were attributed more votes, which made them appear less as outliers. A different victor was sometimes also selected. However, the essential discrepancies from real behavioral observations persisted. The effect of CoT reasoning was inconsistent between hotspots. In one instance, GPT-5.2 predicted a few clicks in areas ignored by GPT-4.1, which indicated the misunderstanding of labels also present in human data. In other instances, GPT-5.2 doubled down on its fidelity errors. The model also overcorrected, either by further emphasizing the victors or by creating new misalignments for hotspots where GPT-4.1 correctly assumed few or zero clicks.

As a result for \hyperref[sec:rq3]{RQ3}, we conclude that CoT reasoning does not moderate the fidelity of LLM-generated first clicks in a meaningful manner. It can also further reduce the diversity of simulated open-ended feedback.

\subsection{Temperature sampling moderation (RQ4)}

\emph{\hyperref[sec:rq4]{RQ4}: How does temperature sampling affect the accuracy of LLM-generated first clicks?}\\

Reducing the \texttt{temperature} to 0.2 had no significant effect on the behavioral fidelity of simulated clicks and the reasoning fidelity as gauged though close-ended questions (see \autoref{fig:click-measures}). In open-ended questions as shown in \autoref{fig:open-boxplots}, semantic diversity was the singular measure where the higher \texttt{temperature} of the baseline led to a better value compared to its reduction, \textit{M} = .66, \textit{SD} = .14 ($z = 2.90$, $p = .004$, $r = .31$ $[.10, .49]$).

As such, the temperature sampling has only a limited capacity to moderate the fidelity synthetic results (\hyperref[sec:rq4]{RQ4}). Essentially, higher temperature can make open-ended answers slightly more semantically diverse without affecting deeper inconsistencies between synthetic and real data.

\subsection{Nucleus sampling moderation (RQ5)}

\emph{\hyperref[sec:rq5]{RQ5}: How does nucleus sampling affect the accuracy of LLM-generated first clicks?}\\

As seen in \autoref{fig:click-measures}, the results for nucleus sampling (\texttt{top\_p}) were similar to the changed \texttt{temperature} (\hyperref[sec:rq4]{RQ4}). Reducing the \texttt{top\_p} to 0.2 yielded no significant differences in measures describing click behavior or close-ended answers. In open-ended answers, there were two significant differences against the baseline shown in \autoref{fig:open-boxplots}. Lower semantic diversity increased the distance from real data \textit{M} = .65, \textit{SD} = .14 ($z = 3.09$, $p = .002$, $r = .33$ $[.13, .51]$) whereas higher readability reduced it \textit{M} = 60.60, \textit{SD} = 21.44 ($z = -2.17$, $p = .031$, $r = .23$ $[-.42, -.02]$).

To answer \hyperref[sec:rq5]{RQ5}, the capacity of nucleus sampling to moderate the fidelity of synthetic data that captures click behavior and the related reasoning is limited. As the result of broader word sampling, the baseline’s higher \texttt{top\_p} ensures slightly more semantically diverse sentences that are less readable.

\subsection{Persona cardinality moderation (RQ6)}

\emph{\hyperref[sec:rq6]{RQ6}: How does persona cardinality (individual vs. population) affect the accuracy of LLM-generated first clicks?}\\

In most aspects, single-persona simulations diverged from the real results significantly more than the mega-persona baseline (see \autoref{fig:click-measures}). Simulation of participants individually resulted in significantly greater Distribution Difference in 96\% of the tasks ($\chi^2=18.92$, $p < .001$, $V=.46$ $[.32, .63]$). Jensen-Shannon Distance was also significantly larger \textit{M} = .58, \textit{SD} = .21 ($z=-2.16$, $p = .031$, $r = -.23$ $[-.42, -.03]$). The reason for this high inconsistency with the ground truth lies in extremely low entropy \textit{M} = .02, \textit{SD} = .06, ($z = 7.97$, $p < .001$, $r = .84$ $[.81, .86]$) as the Unique Selection Count indicates a uniform focus on just a single option despite simulating many different personas \textit{M}=1.16, \textit{SD} = 0.37 ($z = 8.14$, $p < .001$, $r = .86$ $[.85, .86]$). An identical phenomenon was observed in close-ended answers (see \autoref{tab:stats}). Alignment with humans was observed only in a small number of tasks and questions where humans also reached near-absolute consent.

In open-ended questions (see \autoref{fig:similarities} and \autoref{fig:open-boxplots}), single-persona responses had significantly lower semantic similarity to real answers \textit{M} = .61, \textit{SD} = .20 ($z = 2.17$, $p = .03$, $r = .23$ $[.03, .42]$) and semantic diversity \textit{M} = .30, \textit{SD} = .14 ($z = 7.55$, $p < .001$, $r = .80$ $[.71, .85]$). They were exceptionally more repetitive \textit{M} = 7655, \textit{SD} = 3183 ($z = -6.83$, $p < .001$, $r = -.72$ $[-.83, -.59]$) and more verbose \textit{M} = 33.72, \textit{SD} = 17.24 ($z = -6.63$, $p < .001$, $r = -.70$ $[-.82, -.55]$).

Therefore regarding \hyperref[sec:rq6]{RQ6}, persona cardinality moderates the algorithmic fidelity of LLM-generated first click behavior and reasoning. More specifically, the use of a mega-persona over single-personas contributes to higher algorithmic fidelity, although this is mostly indicative of the poor results of single-personas, considering the limitations of mega-personas (see the analysis of \hyperref[sec:rq1]{RQ1} and \hyperref[sec:rq2]{RQ2}).

\subsection{Persona specificity moderation (RQ7)}

\emph{\hyperref[sec:rq7]{RQ7}: How does persona specificity (generic vs. real sample definition) affect the accuracy of LLM-generated first clicks?}\\

There were no significant differences in simulated click behavior between the baseline, where the personas were differentiated to simulate the varied characteristics of the real samples, and the nondescript personas of the alternative condition. Similar observations occurred within open-ended questions as none of the text measures indicated significant differences. The only real difference was confirmed in the entropy of close-ended follow-up answers, which was \textit{M} = .71, \textit{SD} = .14 ($z = 2.10$, $p = .036$, $r = .17$ $[.01, .33]$) for generic personas — slightly lower than the baseline and actually closer to the real data.

The above observations lead to the following answer to \hyperref[sec:rq7]{RQ7}. Persona specificity implemented in the baseline does not significantly affect the accuracy of LLM-simulated first click results, and may even reduce statistical alignment with real data for some metrics.

\section{Discussion}
\label{sec:discussion}

\subsection{Theoretical Implications}

Compared to other behavioral usability research methods, FCT should be simulation-friendly on account of its simplified nature. In the real world, complex human behaviors (e.g., browsing a website, socializing, planning a road trip) are displayed as hierarchical multimodal interactions with the environment observed through continuous sensory inputs. FCT exercises stricter controls by exposing the participant (or agent) to a static stimulus — an image — while also constraining them to a prescriptive atomic reaction — a single click. At the same time, the method maintains its practical grounding by operating with real designs and user tasks. Therefore, FCTs present a suitable experimental milestone that agentic simulators (LLM-driven or otherwise) should be able to pass before attempting to simulate more advanced human behaviors in interactive digital environments. 

Our study provides clear empirical evidence of misalignment between generated outputs and real humans’ cognitive processes, decision-making and behaviors. Our findings, synthesized from patterns across a multitude of diverse studies extend the depth and generalizability of  the current knowledge about the lack of nuance and realism in usability testing performed by LLMs \citep{xiang2024, lu2025}. The inability of the models to click and respond to the GUI designs in a way that would reflect personal individuality and real-world complexity (cognitive, social, epistemic, skill-based, experiential, aesthetic, and procedural factors) expands the understanding of their failure modes in making human-like inferences and interpretations based on sensory stimuli \citep{imschloss2025}.

From the theoretical perspective of cognitive psychology, the differences between real and simulated first click behavior can be attributed to LLMs’ lack of cognitive processes, including attention, genuine intuitive and reflective reasoning, memory and affective cognition. Similar findings are supported by prior experiments in controlled environments, such as economic games demonstrating differences from human decision-making \citep{gao2025}, and math puzzles that showed the failure of chain-of-thought to replicate human-like reasoning \citep{shojaee2025}. Multiple behavioral and self-report patterns we identified in the context of practical first click decision-making — overreliance on victors, leaps in logic, spurious connections — can be reasonably seen to stem from semantic heuristics based on co-occurrences of words in text, which are fundamentally different from mental processing and representation of tasks.

The presence of victors can be interpreted as a form of hyperaccuracy \citep{amirova2024}. This is the known tendency of LLMs to provide uniformly “correct” responses to prompts, typically to provide factual responses to questions. We can speculate that LLM interprets first click tasks as having objectively correct solutions. Among close-ended questions, attitudinal questions were not subject to hyperaccuracy (arguably due to their subjective nature) while those with an assumed “correct” solution were (e.g., “In the image you saw, did you notice [essential information]?” which skewed toward “Yes”).

Linguistics as the foundation of natural language processing presents another valuable explanative lens. LLMs can effectively capture the meaning of sentences through semantic encoding. However, above the literal semantic meaning, pragmatics form a linguistic layer that captures the use and intent of communication \citep{bach2006}. Linguistically, the issues manifested with synthetic first click behavior and reasoning are linked to their literalistic interpretation. LLMs’ constraint to semantics without pragmatics, also reported by \citet{solidjonov2026}, suggests a flawed capacity for pragmatic inference and reasoning (e.g., for annotation of intents or presuppositions).

However, from the behavioral perspective, a deeper issue resides beyond the inability to interpret pragmatics explicitly, i.e. on a semantic level. It is the failure to apply pragmatics-as-pragmatics, as evidenced by two categories of failure modes in our behavioral simulation. The first category concerns reactions to pragmatics implied within the inputs (e.g., first click prediction hampered by inability to infer nuanced stimulus meaning, over-roleplay exaggerating the tasks, hyperfocus and spurious associations based on marginal details). The second links the lack of subjectivity with the lack of personally expressed pragmatics (e.g., missing cognitive, experiential and hedonic factors in responses). Hallucinations, as well as the loss of implicit feedback and tangents are emblematic of semantics where the pragmatics are only superficially approximated.

The underwhelming performance of LLM simulations extends to the reasoning model GPT-5.2 and multiple types of personas \citep{gerosa2024}. CoT reasoning has minimal effects on synthetic behavior, which aligns with its limited effectiveness in solving complex problems \citep{shojaee2025}. While the term “reasoning” presents a useful metaphor for theoretical study of machine psychology, we propose referring to CoT as “realignment” rather than “reasoning” to avoid triggering misleading expectations. Furthermore, the low fidelity of the single-persona approach contextualizes the superficially more human-aligned results of mega-personas. Simulating populations instead of individuals does not mean the model suddenly develops comprehension of people and their embodied experience. In human populations, varied behaviors are observed between individuals because of the different thought processes. The main difference stems from variability. GPT depends on simulating a group to generate what appears as believable diversity. From the perspective of practical algorithmic fidelity, mega-persona simulation is comparable to the single-persona approach, in spite of their statistical differences associated with semantic volume.

\subsection{Practical Implications}

Our study demonstrates that even in simple FCT behavioral scenarios, practitioners should avoid relying on LLMs as a source of behavioral predictions. More specifically, hypothetical adoption as synthetic feedback can lead to misleading conclusions and failure to identify usability issues. Areas of interest interpreted as solutions by large segments of real samples can be missed or only represented as outliers in synthetic data, therefore not showing errors made by real people. Depending on the alignment between the intended correct solutions and the victors selected by the LLM, design teams would likely be lulled into a false sense of security by inflated correctness, or spend effort on fixing hallucinated issues.

Additionally, reliance on LLM-driven first click simulation can blur the understanding of key nuance. For example, users can have a clear behavioral preference because of a reason that the model ignores due to alternative click solutions that appear similar. This smoothing effect is present in various forms, including follow-up responses in Likert scale questions as shown by overly similar distributions in self-reported reasoning in \autoref{fig:likert-diversity}.

LLM can be expected to heuristically select solutions with semantic associations with the task, especially if its label contains similar language. The results do not provide insights into cognitive processes associated with implied meaning, visual and logical hierarchy, positioning, layout, and individual factors such as expectations and mental representations of tasks. As a result, when there is a clear semantic link between tasks and labels found in the stimulus, the LLM will more accurately predict the design intent than the contextual behavior of actual users. Its reinforcement of confirmation bias contradicts the aims and principles of usability research.

Practitioners should be skeptical toward future iterations of LLM-driven behavioral simulation. The “reasoning” GPT-5.2 model and sample-specific mega-persona prompting showed negligible fidelity improvements limited to boosts in believability (e.g., more elaborate responses and diverse behaviors). This very article could be appropriated as an LLM fine-tuning blueprint to obtain behavioral and reasoning responses in FCT that appear more human-like. Models with a more believable façade could hide some of their symptoms (e.g., by hallucinating past experiences and cognitive processes), although other aspects could defy such efforts (e.g., implicit feedback). However, steering and upscaling to achieve better statistics would be committing to a hyperempiricist trap \citep{guest2025}, which does not address the inherent issues of the architectural paradigm. Semantic extrapolation from contextual information does not grant an ability to understand the real world.

Our thematic comparison of real and simulated FCT results provides actionable information for detection of fake responses to researchers and platforms. Bots that compromise the research integrity on some crowdsourcing platforms (e.g., MTurk) are a well-documented problem, predating the wider adoption of LLMs \citep{kennedy2020}. The patterns discussed above could serve as guidelines for quantitative and qualitative detection methods. Extraction of relevant cues is key for machine learning methods used as automated countermeasures against contamination by LLM simulacra. Practitioners looking to recognize LLM responses in their data should be aware that people can also reply sparsely, not explain themselves, or have a positivity bias. The ease at which LLMs can churn out believable data should be seen as a catalyst, highlighting the key role of research design that enables expression of authentic behavior and deep self-reporting. FCT and other quantitative methods may benefit from nesting qualitative design elements to detect fake as well as real-but-careless participants \citep{wang2023}.

Instead of substitutes, quick heuristic feedback when piloting studies could be proposed as a methodologically sounder use for LLMs in research \citep{lu2025}. While reduced harm is a compelling argument, preliminary feedback about study design (e.g., “Will people understand this question?”) can be expected to face identical fidelity issues. In the long-term, trust in the LLM outputs could result in cognitive offloading, deterioration of design skills \citep{shukla2025}, with studies resembling generic adaptations of methods described in the model's training materials rather than attuned to the human audience within the real research context.

\subsection{Limitations and Future Work}

The high ecological validity of our study — owing to our rigorous analysis of data from twelve diverse FCT studies from real practice — has methodological tradeoffs. Firstly, our dataset comprises protected third-party data, reducing the data transparency and reproducibility of our research. Secondly, without controls for sample and study design variables such as the complexity of the stimuli, their contents, task and question counts, as well as phrasing, our results might be skewed by confounding variables. However, we have not identified any systematic factor within the studies that would recontextualize our key findings aside from potentially adding further nuance. During stages of our reflexive thematic analysis, we reflected on the study context (e.g., the visual complexity and layouts of the stimuli, since the majority of studies targeted desktop websites). The tool used to conduct the original studies (UXtweak First Click Testing) implements the standard framework of the FCT activity.

To address some of the knowledge gaps left by our study, future works could prioritize internal validity over generalizability by controlling and examining the effects of specific variables. Controlling research design standards, such as by excluding leading tasks (which we explicitly kept and discussed in our study for realism) may highlight quantitatively worse algorithmic fidelity when only methodologically sound research is considered. Other salient variables include stimulus type (e.g., dashboard, search result page), cognitive complexity and domain knowledge requirements or question types. RTA using a comparable dataset of real FCT studies may yield further equally valid qualitative insights. Researchers undertaking such a pursuit should avoid reading our findings to mitigate confirmation bias.

Given the purpose of FCT is to explore user behavior, we do not investigate pre-training or prompting models with real data as methods for improving model alignment. \citet{park2026} demonstrated in surveys that digital twins created with survey data and transcripts answered questions comparably to demographic and descriptive persona baselines when their answers were not directly retrievable or inferable from the data they were provided with. While not certain without a replication study, it is reasonable to assume a similar effect would be observed if digital twins of participants were created with FCT task data and then evaluated on different FCT tasks completed by the same person.

Our reflexive thematic analysis, regardless of its inductive focus, uses an exploratory framework that leverages the subjective theoretical understanding of the researcher to interpret highly varied data with contextual nuance \citep{braun2020}. For transparency, we declare that the identified themes were informed by the current theory about the algorithmic fidelity of LLM-simulated research participants \citep{kuric2026slr2, kuric2026cssim}, automated detection of usability issues, from traditional rule-based systems to machine learning \citep{kuric2026slr1},  general LLMs capabilities and their inherent limitations (e.g., hallucinations \citep{zhang2026}), AI agents and deep learning applied to UX research contexts \citep{kuric2026talksurv, kuric2025et}, as well as our broader understanding of user behavior and UX research methods \citep{kuric2025cs1, kuric2025hotspots, kuric2025tt}.

\section{Conclusion}
\label{sec:conclusion}

Among the use cases into which LLMs have been inaptly integrated while chasing the lure of maximizable productivity, synthetic participants are particularly untenable. The scientific consensus maintains that LLMs — being probabilistic models trained on multimodal online data rather than capturing actual latent thoughts and contextually nuanced behaviors — cannot substitute observation of real humans. The aim of our study was to obtain a profound understanding of the effects that using LLMs for synthetic behavioral insights can have in real practice. We achieved this by conducting a robust evaluation of twelve real-world study simulations, focusing on the practical and intuitive method First Click Testing (FCT). 

Our analysis revealed a multitude of high-severity problems, including misleading simulated clicks and reasoning caused by the mismatch between the models’ semantic associations and human cognitive processing. Other symptoms of LLM responses included the lack of implicit feedback, exaggerated positivity and helpfulness, hallucinations, along with a literalistic interpretation of language that struggled with pragmatics. Additionally, controlled simulation of LLM parameters, defined personas and CoT reasoning (or “realignment” due to its incongruity with real reasoning) had only superficial effects. Since most identified pattern can be attributed to the mechanics by which LLMs generate text, the limitations of LLM-driven synthetic participants can be expected to apply prospectively, unless they are explicitly and comprehensively addressed by future models.

We propose that the empirical contributions of this study are socially beneficial, supporting research driven decision-making in multiple areas. Our primary intent is informed prevention of cognitive offloading and skill degradation among UX design and research practitioners who may overrely on AI \citep{shukla2025}. The provided evidence may also help secure buy-in by stakeholders from non-research backgrounds, fostering methodologically sounder research and better processes within organizations. The identified issues in simulated first click behavior and reasoning may aid procurement specialists in creating more rigorous verification criteria in face of aggressive AI tool marketing. Additionally, they may serve as the basis for identification and automated detection of LLM-generated fake responses.

\section*{Funding}
This work was supported by the EU NextGenerationEU through the Recovery and Resilience Plan for Slovakia under the project No. 17I04-04-V05-00029, and co-financed by the Slovak Research and Development Agency under Contract No. APVV-23-0408.

\section*{Acknowledgements}
We would like to thank the UXtweak Research team for their technical and expert support.

\section*{Data availability statement}
\label{sec:data-statement}
Supplementary materials (such as LLM prompts utilized in this study) are available in the paper repository at \url{https://github.com/uxtweak-research/synthetic-first-click}.

\section*{Declaration of competing interests}
The authors declare that they have no known competing financial interests or personal relationships that could have appeared to influence the work reported in this paper.

\section*{CRediT authorship contribution statement}
\emph{Eduard Kuric:} Writing – review \& editing, Writing – original draft, Validation, Supervision, Resources, Project administration, Methodology, Investigation, Funding acquisition, Formal analysis, Conceptualization;
\emph{Peter Demcak}: Writing – review \& editing, Writing – original draft, Visualization, Validation, Methodology, Investigation, Formal analysis;
\emph{Matus Krajcovic}: Writing – review \& editing, Writing – original draft, Visualization, Validation, Software, Methodology, Investigation, Formal analysis, Data curation.

\setlength{\bibsep}{0pt}
\bibliography{sources}

\appendix

\section{Extended statistical results}\label{app:stats}

For extended statistical results, see \autoref{tab:stats}.

\begin{sidewaystable}[!ht]
\makeatletter
\renewcommand{\@makecaption}[2]{%
  {\sffamily\footnotesize\bfseries #1\par}
  \vspace{1pt}
  {\sffamily\footnotesize #2\par}
  \vspace{2pt}
}
\makeatother
\caption{Statistical analysis results, comparing simulations to real data and to the GPT-4.1 baseline model. Statistical significance is indicated by asterisks: *p < .05, **p < .01, and ***p < .001.}
\label{tab:stats}
\resizebox{\textheight}{!}{
\begin{tabular}{llllllll}
\toprule
Variable & Real & Baseline & GPT-5.2 (medium reasoning) & Lower temperature & Lower top\_p & Single-persona & Nondescript personas \\
\midrule
\textit{Tasks}  & & & & & & & \\ 
\midrule
Distribution Difference &  & 53\% & \begin{tabular}[t]{@{}l@{}}44\% \\ $\chi^2$=0.40, V=.07\end{tabular} & \begin{tabular}[t]{@{}l@{}}53\% \\ $\chi^2$=0, V=0\end{tabular} & \begin{tabular}[t]{@{}l@{}}62\% \\ $\chi^2$=0.41, V=.07\end{tabular} & \begin{tabular}[t]{@{}l@{}}96\% \\ $\chi^2$=18.92***, V=.46\end{tabular} & \begin{tabular}[t]{@{}l@{}}53\% \\ $\chi^2$=0, V=0\end{tabular} \\ \addlinespace[2pt]
First-choice Agreement &  & 71\% & \begin{tabular}[t]{@{}l@{}}69\% \\ $\chi^2$=0, V=0\end{tabular} & \begin{tabular}[t]{@{}l@{}}71\% \\ $\chi^2$=0, V=0\end{tabular} & \begin{tabular}[t]{@{}l@{}}71\% \\ $\chi^2$=0, V=0\end{tabular} & \begin{tabular}[t]{@{}l@{}}76\% \\ $\chi^2$=0.06, V=.03\end{tabular} & \begin{tabular}[t]{@{}l@{}}71\% \\ $\chi^2$=0, V=0\end{tabular} \\ \addlinespace[2pt]
Rank-order Agreement &  & .16 ± .14 & \begin{tabular}[t]{@{}l@{}}.19 ± .16 \\ z=-0.67, r=-.07\end{tabular} & \begin{tabular}[t]{@{}l@{}}.15 ± .14 \\ z=0.08, r=.01\end{tabular} & \begin{tabular}[t]{@{}l@{}}.14 ± .14 \\ z=0.49, r=.05\end{tabular} & \begin{tabular}[t]{@{}l@{}}.11 ± .08 \\ z=1.15, r=.12\end{tabular} & \begin{tabular}[t]{@{}l@{}}.15 ± .14 \\ z=0.18, r=.02\end{tabular} \\ \addlinespace[2pt]
Jensen-Shannon Distance &  & .48 ± .16 & \begin{tabular}[t]{@{}l@{}}.45 ± .16 \\ z=0.65, r=.07\end{tabular} & \begin{tabular}[t]{@{}l@{}}.48 ± .16 \\ z=0.16, r=.02\end{tabular} & \begin{tabular}[t]{@{}l@{}}.48 ± .16 \\ z=-0.11, r=-.01\end{tabular} & \begin{tabular}[t]{@{}l@{}}.58 ± .21 \\ z=-2.16*, r=-.23\end{tabular} & \begin{tabular}[t]{@{}l@{}}.48 ± .17 \\ z=-0.07, r=-.01\end{tabular} \\ \addlinespace[2pt]
Entropy & .61 ± .18 & .43 ± .19 & \begin{tabular}[t]{@{}l@{}}.54 ± .14 \\ z=-2.77**, r=-.29\end{tabular} & \begin{tabular}[t]{@{}l@{}}.45 ± .19 \\ z=-0.6, r=-.06\end{tabular} & \begin{tabular}[t]{@{}l@{}}.45 ± .19 \\ z=-0.53, r=-.06\end{tabular} & \begin{tabular}[t]{@{}l@{}}.02 ± .06 \\ z=7.97***, r=.84\end{tabular} & \begin{tabular}[t]{@{}l@{}}.45 ± .19 \\ z=-0.33, r=-.03\end{tabular} \\ \addlinespace[2pt]
Entropy gap (vs. real) &  & $\Delta$=-.18, z=-4.09*** & $\Delta$=-.05, z=-1.63 & $\Delta$=-.16, z=-3.80*** & $\Delta$=-.15, z=-3.57*** & $\Delta$=-.63, z=-8.16*** & $\Delta$=-.16, z=-3.77*** \\ \addlinespace[2pt]
Unique Selection Count & 9.00 ± 4.75 & 5.38 ± 2.44 & \begin{tabular}[t]{@{}l@{}}6.51 ± 1.93 \\ z=-3.49***, r=-.37\end{tabular} & \begin{tabular}[t]{@{}l@{}}5.31 ± 2.50 \\ z=0.31, r=.03\end{tabular} & \begin{tabular}[t]{@{}l@{}}5.60 ± 2.72 \\ z=-0.21, r=-.02\end{tabular} & \begin{tabular}[t]{@{}l@{}}1.16 ± 0.37 \\ z=8.14***, r=.86\end{tabular} & \begin{tabular}[t]{@{}l@{}}5.4 ± 2.84 \\ z=0.56, r=.06\end{tabular} \\ \addlinespace[2pt]
Count gap (vs. real) &  & $\Delta$=-3.62, z=-4.35*** & $\Delta$=-2.49, z=-2.22* & $\Delta$=-3.69, z=-4.54*** & $\Delta$=-3.40, z=-3.97*** & $\Delta$=-7.84, z=-8.17*** & $\Delta$=-3.60, z=-4.49*** \\ 
\midrule
\textit{Close-ended questions} & & & & & & & \\ 
\midrule
Distribution Difference &  & 45\% & \begin{tabular}[t]{@{}l@{}}62\% \\ $\chi^2$=3.71, V=.15\end{tabular} & \begin{tabular}[t]{@{}l@{}}45\% \\ $\chi^2$=0, V=0\end{tabular} & \begin{tabular}[t]{@{}l@{}}46\% \\ $\chi^2$=0, V=0\end{tabular} & \begin{tabular}[t]{@{}l@{}}85\% \\ $\chi^2$=25.27***, V=.40\end{tabular} & \begin{tabular}[t]{@{}l@{}}41\% \\ $\chi^2$=0.10, V=.03\end{tabular} \\ \addlinespace[2pt]
First-choice Agreement &  & 50\% & \begin{tabular}[t]{@{}l@{}}38\% \\ $\chi^2$=1.66, V=.10\end{tabular} & \begin{tabular}[t]{@{}l@{}}51\% \\ $\chi^2$=0, V=0\end{tabular} & \begin{tabular}[t]{@{}l@{}}64\% \\ $\chi^2$=2.62, V=.13\end{tabular} & \begin{tabular}[t]{@{}l@{}}73\% \\ $\chi^2$=7.83**, V=.22\end{tabular} & \begin{tabular}[t]{@{}l@{}}62\% \\ $\chi^2$=1.66, V=.10\end{tabular} \\ \addlinespace[2pt]
Rank-order Agreement &  & .36 ± .27 & \begin{tabular}[t]{@{}l@{}}.30 ± .29 \\ z=1.72, r=.14\end{tabular} & \begin{tabular}[t]{@{}l@{}}.39 ± .31 \\ z=-0.44, r=-.04\end{tabular} & \begin{tabular}[t]{@{}l@{}}.40 ± .30 \\ z=-0.84, r=-.07\end{tabular} & \begin{tabular}[t]{@{}l@{}}.33 ± .28 \\ z=0.82, r=.07\end{tabular} & \begin{tabular}[t]{@{}l@{}}.39 ± .32 \\ z=-0.42, r=-.03\end{tabular} \\ \addlinespace[2pt]
Jensen-Shannon Distance &  & .28 ± .13 & \begin{tabular}[t]{@{}l@{}}.30 ± .14 \\ z=-1.74, r=-.14\end{tabular} & \begin{tabular}[t]{@{}l@{}}.28 ± .15 \\ z=-0.18, r=-.01\end{tabular} & \begin{tabular}[t]{@{}l@{}}.27 ± .15 \\ z=0.61, r=.05\end{tabular} & \begin{tabular}[t]{@{}l@{}}.43 ± .19 \\ z=-5.66***, r=-.45\end{tabular} & \begin{tabular}[t]{@{}l@{}}.26 ± .15 \\ z=1.01, r=.08\end{tabular} \\ \addlinespace[2pt]
Entropy & .72 ± .15 & .76 ± .08 & \begin{tabular}[t]{@{}l@{}}.77 ± .11 \\ z=-1.52, r=-.12\end{tabular} & \begin{tabular}[t]{@{}l@{}}.76 ± .09 \\ z=0.21, r=.02\end{tabular} & \begin{tabular}[t]{@{}l@{}}.71 ± .15 \\ z=1.01, r=.08\end{tabular} & \begin{tabular}[t]{@{}l@{}}.34 ± .25 \\ z=9.64***, r=.77\end{tabular} & \begin{tabular}[t]{@{}l@{}}.71 ± .14 \\ z=2.1*, r=.17\end{tabular} \\ \addlinespace[2pt]
Entropy gap (vs. real) &  & $\Delta$=.04, z=0.75 & $\Delta$=.07, z=3.22** & $\Delta$=.04, z=1.03 & $\Delta$=-.01, z=-0.15 & $\Delta$=-.43, z=-9.2*** & $\Delta$=-.01, z=-0.77 \\ \addlinespace[2pt]
Unique Selection Count & 5.15 ± 1.67 & 4.86 ± 1.31 & \begin{tabular}[t]{@{}l@{}}5.5 ± 1.55 \\ z=-3.17***, r=-.25\end{tabular} & \begin{tabular}[t]{@{}l@{}}4.69 ± 1.32 \\ z=1.44, r=.12\end{tabular} & \begin{tabular}[t]{@{}l@{}}4.58 ± 1.41 \\ z=1.67, r=.13\end{tabular} & \begin{tabular}[t]{@{}l@{}}2.47 ± 1.00 \\ z=8.89***, r=.71\end{tabular} & \begin{tabular}[t]{@{}l@{}}4.63 ± 1.63 \\ z=1.65, r=.13\end{tabular} \\ \addlinespace[2pt]
Count gap (vs. real) &  & $\Delta$=-0.29, z=-1.62 & $\Delta$=0.35, z=1.24 & $\Delta$=-0.46, z=-2.19* & $\Delta$=-0.57, z=-2.49* & $\Delta$=-2.68, z=-8.44*** & $\Delta$=-0.52, z=-2.01*\\
\midrule
\textit{Open-ended questions}  & & & & & & & \\ 
\midrule
Lexical Similarity &  & .21 ± .11 & \begin{tabular}[t]{@{}l@{}}.22 ± .10\\ z=-0.24, r=-.03\end{tabular} & \begin{tabular}[t]{@{}l@{}}.21 ± .11\\ z=0.21, r=.02\end{tabular} & \begin{tabular}[t]{@{}l@{}}.20 ± .11\\ z=0.46, r=.05\end{tabular} & \begin{tabular}[t]{@{}l@{}}.22 ± .11\\ z=-0.17, r=-.02\end{tabular} & \begin{tabular}[t]{@{}l@{}}.21 ± .1\\ z=0.07, r=.01\end{tabular} \\ \addlinespace[2pt]
Semantic Similarity &  & .7 ± .15 & \begin{tabular}[t]{@{}l@{}}.68 ± .19\\ z=0.08, r=.01\end{tabular} & \begin{tabular}[t]{@{}l@{}}.68 ± .15\\ z=0.95, r=.10\end{tabular} & \begin{tabular}[t]{@{}l@{}}.67 ± .16\\ z=1.16, r=.12\end{tabular} & \begin{tabular}[t]{@{}l@{}}.61 ± .20\\ z=2.17*, r=.23\end{tabular} & \begin{tabular}[t]{@{}l@{}}.71 ± .15\\ z=-0.24, r=-.03\end{tabular} \\ \addlinespace[2pt]
Semantic Diversity & .76 ± .09 & .7 ± .14 & \begin{tabular}[t]{@{}l@{}}.64 ± .14\\ z=3.0**, r=.32\end{tabular} & \begin{tabular}[t]{@{}l@{}}.66 ± .14\\ z=2.9**, r=.31\end{tabular} & \begin{tabular}[t]{@{}l@{}}.65 ± .14\\ z=3.09**, r=.33\end{tabular} & \begin{tabular}[t]{@{}l@{}}.30 ± .14\\ z=7.55***, r=.80\end{tabular} & \begin{tabular}[t]{@{}l@{}}.69 ± .13\\ z=0.83, r=.09\end{tabular} \\ \addlinespace[2pt]
Semantic Diversity gap (vs. real) &  & $\Delta$=-.06, z=-2.54* & $\Delta$=-.12, z=-4.93*** & $\Delta$=-.10, z=-4.51*** & $\Delta$=-.11, z=-4.77*** & $\Delta$=-.46, z=-8.07*** & $\Delta$=-.07, z=-3.39*** \\ \addlinespace[2pt]
Lexical Repetitiveness & 5606.37 ± 3943.40 & 1579.30 ± 1652.30 & \begin{tabular}[t]{@{}l@{}}4486.25 ± 2603.92\\ z=-5.43***, r=-.57\end{tabular} & \begin{tabular}[t]{@{}l@{}}1770.63 ± 2269.06\\ z=0.03, r=0\end{tabular} & \begin{tabular}[t]{@{}l@{}}2139.92 ± 2255.83\\ z=-0.71, r=-.07\end{tabular} & \begin{tabular}[t]{@{}l@{}}7655.14 ± 3183.16\\ z=-6.83***, r=-.72\end{tabular} & \begin{tabular}[t]{@{}l@{}}1529.34 ± 1524.90\\ z=0.13, r=.01\end{tabular} \\ \addlinespace[2pt]
Lexical Repetitiveness gap (vs. real) &  & $\Delta$=-4027.07, z=-5.62*** & $\Delta$=-1120.12, z=-1.22 & $\Delta$=-3835.74, z=-5.36*** & $\Delta$=-3466.45, z=-4.67*** & $\Delta$=2048.77, z=3.04** & $\Delta$=-4077.03, z=-5.69*** \\ \addlinespace[2pt]
Readability & 67.40 ± 16.27 & 54.04 ± 20.76 & \begin{tabular}[t]{@{}l@{}}53.07 ± 21.66\\ z=0.13, r=.01\end{tabular} & \begin{tabular}[t]{@{}l@{}}59.21 ± 20.84\\ z=-1.84, r=-.19\end{tabular} & \begin{tabular}[t]{@{}l@{}}60.60 ± 21.44\\ z=-2.17*, r=-.23\end{tabular} & \begin{tabular}[t]{@{}l@{}}49.42 ± 23.92\\ z=1.03, r=.11\end{tabular} & \begin{tabular}[t]{@{}l@{}}55.99 ± 18.69\\ z=-0.15, r=-.02\end{tabular} \\ \addlinespace[2pt]
Readability gap (vs. real) &  & $\Delta$=-13.36, z=-3.13** & $\Delta$=-14.33, z=-3.05** & $\Delta$=-8.19, z=-1.62 & $\Delta$=-6.80, z=-1.17 & $\Delta$=-17.98, z=-3.86*** & $\Delta$=-11.41, z=-3.10** \\ \addlinespace[2pt]
Word Count & 13.31 ± 8.26 & 9.08 ± 4.76 & \begin{tabular}[t]{@{}l@{}}13.74 ± 6.59\\ z=-4.80***, r=-.51\end{tabular} & \begin{tabular}[t]{@{}l@{}}8.31 ± 4.09\\ z=1.24, r=.13\end{tabular} & \begin{tabular}[t]{@{}l@{}}8.20 ± 3.55\\ z=0.68, r=.07\end{tabular} & \begin{tabular}[t]{@{}l@{}}33.72 ± 17.24\\ z=-6.63***, r=-.70\end{tabular} & \begin{tabular}[t]{@{}l@{}}8.62 ± 3.50\\ z=-0.10, r=-.01\end{tabular} \\ \addlinespace[2pt]
Word Count gap (vs. real) &  & $\Delta$=-4.23, z=-1.92 & $\Delta$=0.43, z=0.44 & $\Delta$=-5.00, z=-2.51* & $\Delta$=-5.11, z=-2.46* & $\Delta$=20.41, z=6.19*** & $\Delta$=-4.69, z=-2.07*\\
\bottomrule
\end{tabular}%
}
\end{sidewaystable}

\end{document}